\documentstyle[aps,prb,epsf,twocolumn]{revtex}

\catcode`\@=11
\def\simle{\mathrel{\mathpalette\@versim<}}   
\def\simge{\mathrel{\mathpalette\@versim>}}   
\def\@versim#1#2{\lower2.5pt\vbox{\baselineskip0pt \lineskip-.5pt
   \ialign{$\m@th#1\hfil##\hfil$\crcr#2\crcr\sim\crcr}}}

\begin{document}

\draft
\twocolumn[\hsize\textwidth\columnwidth\hsize\csname @twocolumnfalse\endcsname

\title{
Effects of Boson Dispersion in Fermion-Boson Coupled Systems
}
\author{Yukitoshi Motome}
\address{
Institute of Materials Science, University of Tsukuba,
Tsukuba, Ibaraki 305-0006
}
\author{Gabriel Kotliar}
\address{
Department of Physics, Rutgers University,
Piscataway, NJ 08854-8019
}
\date{
\today
}
\maketitle

\begin{abstract}
We study the nonlinear feedback in a fermion-boson system
using an extension of dynamical mean-field theory 
and the quantum Monte Carlo method.
In the perturbative regimes (weak-coupling and atomic limits) 
the effective interaction among fermions increases
as the width of the boson dispersion increases.
In the strong coupling regime away from the anti-adiabatic limit,
the effective interaction decreases
as we increase the width of the boson dispersion.
This behavior is closely related with complete softening of the boson field.
We elucidate the parameters that control this nonperturbative region
where fluctuations of the dispersive bosons enhance the delocalization of fermions.
\end{abstract}

]

\section{Introduction}
\label{Sec:Introduction}

Interacting fermion-boson systems are very important
in condensed matter physics and  have been studied intensively.
\cite{Mahan1981}
They are directly  relevant to the description of electron-lattice interaction.
Other problems can be mapped onto interacting
fermions and bosons by means of the Hubbard-Stratonovich transformation.
\cite{Hubbard1959,Stratonovich1957}
While the  problem of a single fermion interacting with a boson field,
i.e., the polaron problem, is well understood, 
\cite{Mahan1981}
a lot less is known about the many-fermion problem
in interaction with a boson field; it is a full interacting
many-body problem which is only tractable analytically  
in the adiabatic 
\cite{Migdal1958}
and the atomic limits.
\cite{Holstein1959a,Holstein1959b,Mahan1981}

In this paper we revisit the interacting  and dispersive
fermion-boson problem using dynamical mean-field (DMT) theory.
\cite{Georges1996}
This method reduces the quantum many-body problem to a quantum  impurity model
obeying a self-consistency condition. This method has been useful
in describing strong coupling problems such as the  Mott transition.
There are several motivations for our work.

First, a  DMF treatment of the bosonic and fermionic degrees
of freedom  taking into account the boson dispersion, requires an
extension of the  DMF equations where the bosonic
propagator degrees of freedom are determined self-consistently.
This  represents a new type of self-consistent DMF equation,
which so far has not been investigated.
These equations are relevant to many problems,
electron-phonon interactions, fermions interacting with spin fluctuations
\cite{Si9905006}
or among themselves via the  long-ranged Coulomb interactions,
\cite{Chitra9903180}
and to the boson-fermion model.
\cite{Robin9808252}

Second, while the Mott transition in the Hubbard model is well understood
using DMF methods, it is interesting to understand how it is modified
by the variation of the frequency of the mode that mediates the interaction,
or how the results are changed by  the electron-phonon interactions.
The approach discussed in this paper is a first step in this direction.

Finally, phonon dispersion effects are relevant to many systems.
The Jahn-Teller or breathing-type phonons, for instance, seen in manganese oxides
should be dispersive due to intersite coupling.
A distortion of a MnO$_{6}$-octahedron affects distortions of the neighbor octahedra,
since the MnO$_{6}$-octahedra share their oxygen atoms
which leads to an  intersite coupling.
This may be relevant to fascinating orderings of lattice and charge in doped manganites.
\cite{Ramirez1996,Chen1997,Mori1998,Frenandez-Diaz1999}

We study   the mutual feedback of fermionic and
bosonic degrees of freedom in   a very simple system of  fermions
interacting with one branch of bosons   at half filling.
However the  methodology  can  be extended
to other problems where similar DMF equations occur such as
electron problems with long-ranged Coulomb interactions and the competition
of magnetic order and the heavy fermion state,
and to the boson-fermion mixture of high temperature superconductivity.

This paper is organized as follows.
In the next section we discuss how  DMF theory needs to be extended 
to fully include the feedback effects through fermion-boson interaction.
Quantum Monte Carlo (QMC) method is introduced to solve the DMF equations
for all parameter regions. We also discuss
some technical points of the  QMC  relevant to this problem.
The formalism is applied to demonstrate effects of boson dispersion
in a wide region of parameters  and the results
are summarized in  Sec. \ref{Sec:Results}.
In Sec. \ref{Sec:Discussions},
we discuss our main result:
The existence of two distinct regimes of the DMF solutions.
In the first regime, the feedback effects increase the fermion-boson coupling.
In the second regime, which is strongly fluctuating,
the boson dispersion accelerates the delocalization of fermions.
Complete softening characterizes the crossover between these regimes.
Section \ref{Sec:Summary} is devoted to summary.

\section{Dynamical Mean-Field Formalism and Hamiltonian}
\label{Sec:Formalism}

In this work, we discuss such feedback effects
caused by the fermion-boson interaction using DMF theory.
DMF theory provides a local view of a many-body problem
in terms of an impurity model which satisfies a self-consistency condition.
\cite{Georges1996}
For general fermion-boson problems with a local interaction,
the local action has the form
\begin{eqnarray}
\nonumber
S_{\rm eff} &=& \int d\tau d\tau'
\sum_{\alpha} c_{\alpha}^{\dagger} (\tau)
{\cal G}_{0 \alpha}^{-1} (\tau-\tau') c_{\alpha} (\tau')
\\ \nonumber
&+& \int d\tau d\tau' \sum_{\nu} x_{\nu} (\tau)
{\cal D}_{0 \nu}^{-1} (\tau-\tau') x_{\nu} (\tau')
\\
&+& \int d\tau \sum_{\alpha_{1} \alpha_{2} \nu}
\lambda_{\alpha_{1} \alpha_{2} \nu}
c_{\alpha_{1}}^{\dagger} (\tau) c_{\alpha_{2}} (\tau) x_{\nu} (\tau),
\label{eq:Seff}
\end{eqnarray}
where ${\cal G}_{0}$ and ${\cal D}_{0}$ are the bare impurity Green's functions
for fermion and boson, respectively, which contain the dynamical information
of the integrated other sites.
Here $c_{\alpha}$ is the fermion annihilation operator and
$x_{\nu}$ is the boson field.
$\lambda_{\alpha_{1} \alpha_{2} \nu}$ denotes the coupling
between fermions and bosons.
The index $\alpha$ ($\nu$) denotes internal degrees of freedom of
fermions (bosons) such as spins or orbitals of electrons (normal modes of phonons).
We do not explicitly write the contribution from fermion interactions
such as the Coulomb interaction
since we focus on the effects of boson dispersions in this paper.
However the action (\ref{eq:Seff}) is quite general
which contains such fermion interactions
through the Hubbard-Stratonovich transformation
\cite{Hubbard1959,Stratonovich1957}
with continuous fields.
Of course, alternatively, one can include additionally
the fermion interactions according to the DMF theory for the Hubbard-type models.
\cite{Georges1996}

The full Green's functions are related to the bare ones by
\begin{eqnarray}
{\cal G}_{\alpha}^{-1} (i\omega_{n}) &=&
{\cal G}_{0\alpha}^{-1} (i\omega_{n}) - \Sigma_{\alpha} (i\omega_{n}),
\label{eq:relG}
\\
{\cal D}_{\nu}^{-1} (i\omega_{n}) &=&
{\cal D}_{0\nu}^{-1} (i\omega_{n}) - \Pi_{\nu} (i\omega_{n}),
\label{eq:relD}
\end{eqnarray}
at each Matsubara frequency $\omega_{n}=(2n+1)\pi/\beta$ for fermion
and $\omega_{n}=2n\pi/\beta$ for boson, respectively ($n$ is an integer).
$\beta$ is the inverse temperature.
$\Sigma$ and $\Pi$ are the self-energy for fermion and boson, respectively.
The Green's functions for both fermion and boson are
determined in a self-consistent way.
This is achieved by the following set of self-consistency conditions,
\begin{eqnarray}
{\cal G_{\alpha}} &=&
\sum_{\mbox{\boldmath $q$}}
\left[ i\omega_{n} + \mu -\epsilon_{\mbox{\boldmath $q$}\alpha}
- \Sigma_{\alpha} (i\omega_{n}) \right]^{-1},
\label{eq:selfG}
\\
{\cal D_{\nu}} &=&
\sum_{\mbox{\boldmath $q$}}
\left[ (i\omega_{n})^{2} - \omega_{\mbox{\boldmath $q$} \nu}^{2}
- \Pi_{\nu} (i\omega_{n}) \right]^{-1},
\label{eq:selfD}
\end{eqnarray}
where $\epsilon_{\mbox{\boldmath $q$} \alpha}$ and $\omega_{\mbox{\boldmath $q$} \nu}$
give the dispersion relations for fermions and bosons
as a function of the wave number $\mbox{\boldmath $q$}$, respectively.
$\mu$ is the chemical potential to control the density of fermions.
Here the bosons are described as harmonic oscillators.
The condition (\ref{eq:selfD}) is modified
according to the property of boson degrees of freedom.

Previous studies of other models have indicated that
the results of the DMF theory can give useful insights into 
three-dimensional systems.
\cite{Georges1996}
We have therefore taken the dispersions
$\epsilon_{\mbox{\boldmath $q$}}$ and $\omega_{\mbox{\boldmath $q$}}$
that correspond to a semicircular density of states, 
see the details in Sec. \ref{SubSec:Model}.
These DMF equations 
are  exact for a  model where the fermions
and the bosons  have random hopping on lattice sites.  

The self-consistency loop is closed as follows:
The effective action (\ref{eq:Seff}) is solved for given bare impurity Green's functions
${\cal G}_{0}$ and ${\cal D}_{0}$
to obtain the full Green's functions ${\cal G}$ and ${\cal D}$.
The self-energy $\Sigma$ and $\Pi$ are calculated
by the relations (\ref{eq:relG}) and (\ref{eq:relD}),
and used to obtain the Green's functions
through the self-consistency conditions (\ref{eq:selfG}) and (\ref{eq:selfD}).
New bare impurity Green's functions
are calculated by the relations (\ref{eq:relG}) and (\ref{eq:relD}) again.
This loop is iterated until all the quantities are converged.
In this way, both fermionic and bosonic dispersions are renormalized
through the fermion-boson interaction,
and the mutual feedback effects are fully included.

The above DMF equations assume that  
no symmetry breaking is present in the system
although the extension to phases with broken symmetry is straightforward.
And they can be derived from
an electron-phonon model:
\begin{equation}
\label{eq:H}
{\cal H}= {\cal H}_{\rm F} + {\cal H}_{\rm B} + {\cal H}_{\rm I},
\end{equation}
where
\begin{eqnarray}
{\cal H}_{\rm F} &=& \sum_{\alpha} \sum_{\mbox{\boldmath $q$}}
\left( \epsilon_{\mbox{\boldmath $q$} \alpha} - \mu \right)
c_{\mbox{\boldmath $q$} \alpha}^{\dagger}
c_{\mbox{\boldmath $q$} \alpha}
+ {\cal H}_{\rm int},
\label{eq:H_F}
\\
{\cal H}_{\rm B} &=& \sum_{\nu} \sum_{\mbox{\boldmath $q$}}
\omega_{\mbox{\boldmath $q$} \nu}
\left( a_{\mbox{\boldmath $q$} \nu}^{\dagger}
a_{\mbox{\boldmath $q$} \nu} + \frac{1}{2} \right),
\label{eq:H_B}
\\
{\cal H}_{\rm I} &=& \sum_{\alpha_{1} \alpha_{2} \nu}
\sum_{\mbox{\boldmath $p$}, \mbox{\boldmath $q$}}
\tilde{\lambda}_{\mbox{\boldmath $q$} \alpha_{1} \alpha_{2} \nu}
c_{\mbox{\boldmath $p$}+{\mbox{\boldmath $q$} \alpha_{1}}}^{\dagger}
c_{\mbox{\boldmath $p$} \alpha_{2}}
\left( a_{\mbox{\boldmath $q$} \nu} +
a_{-\mbox{\boldmath $q$} \nu}^{\dagger} \right),
\label{eq:H_I}
\end{eqnarray}
where $a_{\mbox{\boldmath $q$}}$ is the boson annihilation operator
which is related with the boson field by
$
x_{\mbox{\boldmath $q$}} = (2M \omega_{\mbox{\boldmath $q$}})^{-1/2}
( a_{\mbox{\boldmath $q$}} +
a_{-\mbox{\boldmath $q$}}^{\dagger} ).
$

The model (\ref{eq:H}) has been  intensively studied using DMF methods
in the  limit of zero  boson dispersion, i.e., in the 
Holstein model:
\cite{Holstein1959a,Holstein1959b}
Bosons with the same index $\nu$ have a same frequency (Einstein phonons)
and the fermion-boson coupling is local as
\begin{eqnarray}
{\cal H}_{\rm B} &=& \sum_{\nu} \omega_{0 \nu} \sum_{\mbox{\boldmath $q$}}
\left( a_{\mbox{\boldmath $q$} \nu}^{\dagger}
a_{\mbox{\boldmath $q$} \nu} + \frac{1}{2} \right)
\nonumber
\\
&=& \sum_{\nu} \frac{M_{\nu}}{2} \sum_{i}
\left( \dot{x}_{i \nu}^{2} + \omega_{0 \nu}^{2} x_{i \nu}^{2} \right),
\label{eq:H_B:Holstein}
\\
{\cal H}_{\rm I} &=& \sum_{\alpha_{1} \alpha_{2} \nu} \sum_{i}
\lambda_{\alpha_{1} \alpha_{2} \nu}
c_{i \alpha_{1}}^{\dagger} c_{i \alpha_{2}} x_{i \nu},
\end{eqnarray}
where
the index $i$ denotes a lattice site.
In the ground state, the possibility of charge-ordered or superconducting states
has been intensively discussed for this model.
\cite{Hirsch1982,Hirsch1983,Scaletter1989,Noack1991,Freericks1993}
Above the critical temperatures of these states,
the crossover behavior is observed
from the Fermi liquid with a mass enhancement in the weak coupling region
to the so-called polaron which is a combined object between fermion and boson
in the strong coupling region.
\cite{Mahan1981,Holstein1959a,Holstein1959b,Millis1996}

It is instructive to compare the present framework
with the DMF theory for the problem without the boson dispersions such as the Holstein model.
If bosons have no dispersion, that is, all $\omega_{\mbox{\boldmath $q$}}$
take the same value $\omega_{0}$ independent of $\mbox{\boldmath $q$}$,
Eq. (\ref{eq:selfD}) is rewritten as
\begin{equation}
{\cal D} = \left[ (i\omega_{n})^{2} - \omega_{0}^{2} - \Pi (i\omega_{n}) \right]^{-1}.
\end{equation}
Although the full Green's function ${\cal D}$ contains a feedback effect in the self-energy $\Pi$, 
the bare impurity Green's function ${\cal D}_{0}$ is fixed
at the noninteracting Green's function given by
\begin{equation}
{\cal D}_{0}^{\rm free} = \left[ (i\omega_{n})^{2} - \omega_{0}^{2} \right]^{-1}
\label{eq:D_0_free}
\end{equation}
throughout the self-consistency iterations
when we start from the solution ${\cal D}_{0} = {\cal D}_{0}^{\rm free}$.
This is equivalent to the ordinary DMF theory for the Holstein model
which does not need Eqs. (\ref{eq:relD}) and (\ref{eq:selfD}).
\cite{Freericks1993,Millis1996}
Compared to this, for the cases with finite bosonic dispersions,
the bare impurity Green's functions ${\cal D}_{0}$ is renormalized
from ${\cal D}_{0}^{\rm free}$ in the iterations in our formalism.

The renormalization of ${\cal D}_{0}$ plays a crucial role
because ${\cal D}_{0}$ is related to the effective interaction between fermions.
If we integrate out the boson variables $x$ in the Hamiltonian,
the effective interaction between fermions takes the form
\begin{equation}
\sum_{\alpha_{1} \alpha_{2} \alpha_{3} \alpha_{4}} \int d\tau d\tau'
c_{\alpha_{1}}^{\dagger} (\tau) c_{\alpha_{2}} (\tau)
U_{\rm eff} (\tau-\tau')
c_{\alpha_{3}}^{\dagger} (\tau') c_{\alpha_{4}} (\tau'),
\end{equation}
where\begin{equation}
U_{\rm eff} (\tau) = \lambda^{2} {\cal D}_{0} (\tau).
\label{eq:U_eff}
\end{equation}
In the absence of the boson dispersion,
since ${\cal D}_{0}$ is unchanged through the self-consistency loop as mentioned above, 
the effective interaction (\ref{eq:U_eff}) is also unrenormalized.
On the other hand, ${\cal D}_{0}$ is renormalized in our formalism for finite dispersions,
which means that the effective interaction between fermions is
renormalized by the mutual feedback of the fermion-boson coupling.

There are several techniques to solve the effective impurity problem
with the action (\ref{eq:Seff}).
In this work, we employ QMC method
\cite{Georges1996,Hirsch1986}
because it is an unbiased calculation and
suitable to investigate all the parameter regions beyond perturbative regimes.
In the QMC approach, the imaginary time is discretized into $L$ slices
with the width $\Delta\tau$ ($\Delta\tau = \beta/L$).
Continuous variables $x_{\nu l} = x_{\nu} (\tau_{l})$
($\tau_{l} = l \Delta\tau$, $l=1, 2, \cdot\cdot\cdot, L$)
are randomly updated to $x'_{\nu l}$ with the probability
\begin{equation}
\prod_{\alpha} \frac{\det{\cal G}_{\alpha}}{\det{\cal G}'_{\alpha}}
\frac{\exp \left[ -\Delta\tau B (x'_{\nu l}) \right]}
{\exp \left[ -\Delta\tau B (x_{\nu l}) \right]},
\label{eq:MCweight}
\end{equation}
where
$
B (x_{\nu l}) =  \sum_{j=1}^{L} x_{\nu j} {\cal D}_{0 \nu jl}^{-1} x_{\nu l}
$
with ${\cal D}_{0 \nu jl} = {\cal D}_{0 \nu} (\tau_{j}-\tau_{l})$.
The fermion Green's functions ${\cal G}$ and ${\cal G}'$ are calculated
by the standard algorithm
\cite{Hirsch1986}
for the configurations with $x_{\nu l}$ and $x'_{\nu l}$, respectively.

In actual QMC samplings, we consider both local and global updates
for the continuous fields $x_{\nu l}$.
The local update consists of sequential updates of the fields on each discretized point;
a change from $x_{\nu l}$ to $x'_{\nu l} = x_{\nu l} + r \delta$ is attempted
where $r$ is a random number between $-1$ and $1$ and $\delta$ is a given amplitude.
The global one is a simultaneous movement of all the fields at a same amount $r \delta$.
The latter becomes important especially in the strong coupling region and/or 
at low temperatures where the fields $x$ show some orderings or are nearly ordered.
The update amplitude $\delta$ is chosen to give an appropriate value of
the acceptance ratio which is defined as the ratio of the number of accepted samples
to the total number of trials.

The QMC calculations generally have the negative sign problem;
the MC weight (\ref{eq:MCweight}) can be negative for the general action (\ref{eq:Seff}),
which leads to numerical instability in the QMC measurements.
However, if fermions couple to bosons only in the diagonal form, that is,
the coupling parameter $\lambda_{\alpha_{1} \alpha_{2} \nu}$ is nonzero
only for the case of $\alpha_{1} = \alpha_{2}$,
the MC weight (\ref{eq:MCweight}) becomes positive definite.
\cite{Takegahara1992}
In this case, there is no negative sign problem for all parameters.

There are two sources of errors in the QMC calculations.
One is a systematic error due to the discretization of the imaginary time, and
the other is a statistical error from the random sampling.
The former error is known to be proportional to $(\Delta \tau^{2})$.
Measurement is divided into several blocks to estimate the latter statistical error
by the variance among the blocks.
The size of each error depends on a specific form of models and parameters.

\section{Results}
\label{Sec:Results}

\subsection{Model and Parameters}
\label{SubSec:Model}

We apply the new DMF framework proposed in the previous section
to a case that the general Hamiltonian (\ref{eq:H}) contains
two species of fermions and one branch of bosons.
We set the mass $M=1$.
The model is a straightforward extension of the Holstein model
to include dispersive bosons, whose fermion-boson interaction is given by
\begin{equation}
{\cal H}_{\rm I} = -\lambda \sum_{i \alpha}
\left( c_{i \alpha}^{\dagger} c_{i \alpha} -\frac{1}{2} \right) x_{i},
\label{eq:modelA}
\end{equation}
where the index $\alpha$ takes two values like spin degrees of freedom of electrons.
The interaction is diagonal in the fermion index $\alpha$
so that the QMC does not suffer from the negative sign problem
as mentioned in Sec. \ref{Sec:Formalism}.
We take the coupling parameter $\lambda$ to be positive,
which favors a doubly-occupied or an empty state on each site.
Note that the model has the particle-hole symmetry at $\mu=0$.

The boson dispersion is taken into account through Eq. (\ref{eq:selfD}) in the present framework.
We replace the summations over the wave number $\mbox{\boldmath $q$}$
in Eqs. (\ref{eq:selfG}) and (\ref{eq:selfD}) by the energy integrations as
\begin{eqnarray}
{\cal G} (i\omega_{n})= \int  \frac{D_{\rm F} (\varepsilon) \ d\varepsilon}
{i\omega_{n} + \mu - \varepsilon -\Sigma (i\omega_{n})},
\label{eq:selfG_int}
\\
{\cal D} (i\omega_{n})= \int  \frac{D_{\rm B} (\varepsilon) \ d\varepsilon}
{(i\omega_{n})^{2} - \varepsilon^{2} - \Pi (i\omega_{n})},
\label{eq:selfD_int}
\end{eqnarray}
where $D_{\rm F}$ and $D_{\rm B}$ are
the the density of states for fermion and boson, respectively.
In the following calculations, we assume semicircular density of states as
\begin{eqnarray}
D_{\rm F} (\varepsilon) &=& \frac{2}{\pi W^{2}} \sqrt{W^{2} - \varepsilon^{2}},
\label{eq:DOS_F}
\\
D_{\rm B} (\varepsilon)  &=& \frac{2}{\pi \omega_{1}^{2}}
\sqrt{\omega_{1}^{2} - (\varepsilon- \omega_{0})^{2}},
\label{eq:DOS_B}
\end{eqnarray}
where $W$ is the half-bandwidth of the fermion density of states
which is taken as unity hereafter ($W=1$);
$\omega_{0}$ and $\omega_{1}$ are the center and the half-bandwidth
of the boson density of states, respectively ($\omega_{0} > 0$, $\omega_{0}-\omega_{1} > 0$).
For the semicircular density of states,
the integrations (\ref{eq:selfG_int}) and (\ref{eq:selfD_int}) are performed analytically
\cite{Georges1996}
which give
\begin{eqnarray}
{\cal G} &=& \frac{\zeta - \sqrt{\zeta^{2} - 4 t^{2}}}{2 t^{2}},
\\
{\cal D} &=& \frac{1}{\xi}
\left[ \frac{1}{\xi_{-} + \sqrt{\xi_{-}^{2} - \omega_{1}^{2}}}
+ \frac{1}{\xi_{+} + \sqrt{\xi_{+}^{2} - \omega_{1}^{2}}} \right],
\end{eqnarray}
where $\zeta = i\omega_{n} + \mu$ and
$\xi_{\pm} = \xi \pm \omega_{0}$ with $\xi^{2} = (i\omega_{n})^{2} - \Pi$.

The shape of the boson density of states near the bottom is important
because bosons at the band edge can be easily excited and strongly interact with fermions.
The semicircular density of states (\ref{eq:DOS_B}) has
an $\varepsilon^{1/2}$-singularity which is expected for bosons
with ordinary cosine dispersions in three dimensions.
Therefore we believe that the following results are qualitatively unchanged
in realistic three-dimensional models.
Results would be different for the  two-dimensional
density of states which has  a step discontinuity  at the band edges and results in  
very different DMF solutions.

In the absence of the boson dispersion ($\omega_{1} = 0$),
the model with the interaction (\ref{eq:modelA}) (the ordinary Holstein model)
shows a charge ordering around half-filling ($\mu =0$) and 
superconductivity in doped regions at very low temperatures.
\cite{Hirsch1982,Hirsch1983,Scaletter1989,Noack1991,Freericks1993}
In the following, we examine effects of boson dispersions
in the low temperature region above and around these transition temperatures
at half-filling ($\mu = 0$) assuming no symmetry breaking.
The calculations are mainly performed at $\beta=8$.
We take $\Delta \tau = 1/4$ for which all the measured quantities are converged
to the limit of $\Delta \tau \rightarrow 0$ within the statistical errors.
We have typically run $1,000,000$ MC steps for measurements;
one MC sampling means a set of a sweep of local updates
over the whole discretized points and a global update.
Convergence in the self-consistency loop is usually rapid;
typically 10 iterations are required to converge within the statistical errorbars
when we start from the noninteracting Green's functions.
However in the strong coupling case,
the iteration often suffers from an oscillation between two solutions.
To avoid the oscillation, we make the iteration proceed by mixing the previous solutions.

\subsection{Dispersionless Boson}
\label{SubSec:Dispersionless Boson}

First, we reconsider the limit without the boson dispersion, that is,
$\omega_{1} = 0$. 
In this case,
we use the two parameters $\omega_{0}$ and $U = \lambda^{2}/M\omega_{0}^{2}$
to characterize basic properties of the system.
The first parameter $\omega_{0}$ describes the adiabaticity.
In the adiabatic limit of $\omega_{0} \rightarrow 0$,
the boson fields do not change in the imaginary time, that is, they behave as classical fields.
In the opposite limit of $\omega_{0} \rightarrow \infty$,
the bosons react instantaneously to fermion motions.
Between these two limits, bosons with a finite $\omega_{0}$ mediate
a retarded effective interaction
which is given by $U_{\rm eff}$ in Eq. (\ref{eq:U_eff}).
The second parameter $U$ describes the magnitude of the effective interaction between fermions.
Note that $U = |U_{\rm eff} (\omega_{n}=0)|$ in this dispersionless case,
since the bare impurity Green's function is given by the noninteracting one (\ref{eq:D_0_free}).

For a fixed value of $\omega_{0}$, the system behaves quite differently
in the regions with $U \ll 1$ and $U \gg 1$.
For small values of $U$, fermions are nearly free and
each lattice site is in an empty, a singly-occupied, or a doubly-occupied state
with almost equal probability at half-filling ($\mu = 0$).
If we define the probability $P(x)$ that
the boson field $x$ lies in the interval between $x$ and $x + \Delta x$,
$P(x)$ shows a single broad peak centered at $x=0$.
Compared to this, if $U$ becomes large, fermions strongly interact with each other, and
a combined state between fermion and boson may be formed,
which is called a small polaron.
The polaron consists of double occupancy of fermions
for the model with the interaction (\ref{eq:modelA}) (bipolaron).
Then, the probability $P(x)$ displays a double peak
at $x= \pm \lambda / M \omega_{0}^{2}$
which corresponds to the doubly-occupied and empty states.
Figure \ref{fig:Om0=0.5,Om1=0} shows this behavior
by changing the value of $U$ for the case of $\omega_{0} = 0.5$.
The single peak of the probability $P(x)$ appears for small $U$,
while the double peaks are developed for $U \simge 1$
as shown in Fig. \ref{fig:Om0=0.5,Om1=0}(a).
At the same time, in Fig. \ref{fig:Om0=0.5,Om1=0}(b),
the probability of the double occupancy $P_{\rm D}$ increases
from $1/4$ for the noninteracting case to $1/2$ for the situation
in which the system consists of only empty and doubly-occupied sites.
The self-energy for fermion $\Sigma$ is also enhanced
by the effective interaction between fermions $U$.
Figure \ref{fig:Om0=0.5,Om1=0}(c) shows that
the absolute value of the imaginary part of the self-energy
as a function of Matsubara frequency is strongly enhanced by $U$.
Note that the data for $\omega_{n} > 1/\Delta\tau$ contain no unbiased information.
These clearly indicate the crossover
from the weakly-correlated fermions in the small $U$ region
to the small polarons in the large $U$ region.
\cite{Millis1996}

The similar crossover is found for other values of $\omega_{0}$.
Figures \ref{fig:Om0=2.0,Om1=0} and \ref{fig:Om0=8.0,Om1=0}
show the results for $\omega_{0} = 2$ and $8$, respectively.
The value of $U$ for the crossover, which we call $U^{*}$ hereafter,
depends on the value of $\omega_{0}$.
For example, for the case of $\omega_{0} = 0.5$ in Fig. \ref{fig:Om0=0.5,Om1=0},
the double-peak structure of $P(x)$ appears at $U \sim1$,
on the other hand, it does not appear up to $U \sim 3$ for $\omega_{0} = 8$.
This can be  understood as follows:
In the limit of $\omega_{0} \rightarrow \infty$,
since the effective interaction becomes spontaneous,
$U_{\rm eff}(\tau) = - U \delta(\tau)$,
the model maps onto an attractive Hubbard model
\cite{Micnus1990}
in which the boson field corresponds to the continuous Hubbard-Stratonovich field.
\cite{Hubbard1959,Stratonovich1957}
In the Hubbard model, it is known that the continuous field develops
a double-peak distribution at $U \sim 3$,
which corresponds to the opening of the Hubbard gap in the case of a repulsive interaction.
\cite{Georges1996}
On the other hand, in the opposite limit of $\omega_{0} \rightarrow 0$,
the effective interaction becomes constant in the imaginary time, $U_{\rm eff}(\tau) = -U$.
This case is identical to an attractive Falicov-Kimball model
in the limit of a continuous number of configurations for the static fields.
\cite{Freericks1993}
In the adiabatic limit, fermions are localized at a smaller value of $U$ since
fluctuations of the boson field is smaller in this case
than in the anti-adiabatic limit.
Then, the splitting of the distribution of $x$ should appear at a lower value of $U$.
In the Falicov-Kimball model with a discrete static field,
the critical value of $U$ is estimated to be $1$.
\cite{van Dongen1991,van Dongen1992}
The finite value of $\omega_{0}$ can interpolate these two limits.
Thus, the value of $U^{*}$ may change smoothly
from $U^{*} \sim 3$ in the limit of $\omega_{0} \rightarrow \infty$
to $U^{*} \sim 1$ in the limit of $\omega_{0} \rightarrow 0$.

\subsection{Dispersive Boson: Weak Coupling Limit}
\label{SubSec:Weak Coupling Limit}

Now we discuss the cases with a finite bosonic dispersion; $\omega_{1} \ne 0$.
First, we study the weak coupling limit of $W \gg \omega_{0}$ and $U$
which has been studied by a perturbation theory.
\cite{Migdal1958}

In this region, the finite width of the boson dispersion $\omega_{1}$
enhances the effective interaction between fermions.
Figure \ref{fig:Om0=0.5,lambda=0.2}(a) shows
the bare impurity Green's function for boson ${\cal D}_{0}$
as a function of Matsubara frequency for various values of $\omega_{1}$
for the case of $\omega_{0}=0.5$ and $U=0.16$ ($\lambda=0.2$).
${\cal D}_{0}$
is enhanced by the width of the dispersion $\omega_{1}$,
which indicates that through the relation (\ref{eq:U_eff}),
the effective interaction between fermions $U_{\rm eff}$
is enhanced by $\omega_{1}$.
This enhancement is also observed in
the imaginary part of the fermion self-energy 
as shown in Fig. \ref{fig:Om0=0.5,lambda=0.2}(b).
At the same time, the probability of the double occupancy becomes large
as shown in Fig. \ref{fig:Om0=0.5,lambda=0.2}(c).
These features are similar to those
in Figs. \ref{fig:Om0=0.5,Om1=0}-\ref{fig:Om0=8.0,Om1=0}
when the parameter $U$ increases in the small $U$ region.
These results can be understood using  a perturbative argument
in Sec. \ref{Sec:Discussions}.

\subsection{Dispersive Boson: Atomic Limit}
\label{SubSec:Atomic Limit}

Next, we consider the limit of $W \ll \omega_{0}$ and $U$
which has been studied based on the so-called small-polaron theory.
\cite{Holstein1959a,Holstein1959b,Mahan1981}
In this limit, the coherent band motion of fermions in Eq. (\ref{eq:H_F}) is
a perturbation on other terms of (\ref{eq:H_B}) and (\ref{eq:H_I}).
The small-polaron theory is a perturbative approach from the atomic limit.
The strong interaction between fermions and bosons leads
to the formation of  the small-polaron state as mentioned in the dispersionless case
in Sec. \ref{SubSec:Dispersionless Boson}.

In this region, as the weak coupling case in Sec. \ref{SubSec:Weak Coupling Limit},
the effective interaction between fermions is enhanced
by the finite width of the boson dispersion.
Figure \ref{fig:Om0=8,lambda=24}(a) plots
the bare impurity Green's function for boson at zero Matsubara frequency
for $\omega_{0}=8$ and $U=9$ ($\lambda=24$).
A finite width of the boson dispersion $\omega_{1}$ enhances ${\cal D}_{0} (\omega_{n}=0)$.
${\cal D}_{0}$ shows the largest change at zero frequency
as in Fig. \ref{fig:Om0=0.5,lambda=0.2}(a).
At the same time, the absolute value of the imaginary part of the fermion self-energy
increases as shown in Fig. \ref{fig:Om0=8,lambda=24}(b).
We plot here the data at the smallest Matsubara frequency
to show the behavior clearly.
The double-peak structure of the probability function $P(x)$
shown in Fig. \ref{fig:Om0=8.0,Om1=0}(a) at $\omega_{1} = 0$,
which indicates the formation of the small-polaron state,
does not change for $\omega_{1}$ within statistical errorbars.
This suggests that the finite width of the boson dispersion enhances the effective interaction
while the small-polaron state remains stable.
These features will be discussed based on the small-polaron theory
in Sec. \ref{Sec:Discussions}.

\subsection{Dispersive Boson: Strong Fluctuation Regime}
\label{SubSec:Strong Fluctuation Regime}

Here we go beyond the perturbative regimes studied in
Sec. \ref{SubSec:Weak Coupling Limit} and \ref{SubSec:Atomic Limit}.
We consider the strong coupling case away from the anti-adiabatic limit,
that is, $U > W$ and $\omega_{0} \sim W$.
It is difficult to study this regime by any perturbative and analytical approach
because of strong fluctuations.
Our DMF method including the fluctuation effects is applied to this regime without any difficulty.

Figure \ref{fig:Om0=0.5,lambda=0.8} shows the results
for $\omega_{0} = 0.5$ and $U = 2.56$ ($\lambda = 0.8$).
As shown in Fig. \ref{fig:Om0=0.5,lambda=0.8}(a),
the absolute value of the bare impurity Green's function for boson ${\cal D}_{0}$
decreases as $\omega_{1}$ increases.
The imaginary part of the self-energy for fermion also decreases its absolute value
as shown in Fig. \ref{fig:Om0=0.5,lambda=0.8}(b).
At the same time, the probability of the double occupancy $P_{\rm D}$ decreases from $1/2$
as shown in Fig. \ref{fig:Om0=0.5,lambda=0.8}(c).
Figure \ref{fig:Om0=0.5,lambda=0.8}(d) shows that
the double-peak structure of the probability $P(x)$ becomes unclear to merge into a single peak.
All these features exhibit that
the effective interaction between fermions $U_{\rm eff}$ is weakened and
the small-polaron state becomes unstable for $\omega_{1}$.
This is a striking contrast to the previous results
in Sec. \ref{SubSec:Weak Coupling Limit} and \ref{SubSec:Atomic Limit}.
We will discuss a physical picture for this behavior in Sec. \ref{Sec:Discussions}.

In the intermediate region,
we find a crossover as the value of $\omega_{1}$ increases.
Figure \ref{fig:Om0=0.5,lambda=0.4} shows this crossover
for $\omega_{0}=0.5$ and $U = 0.64$ ($\lambda = 0.4$).
For small values of $\omega_{1}$,
we find a similar behavior as seen in Fig. \ref{fig:Om0=0.5,lambda=0.2};
the bare impurity Green's function for boson is enhanced and
both the absolute value of the self-energy and the double occupancy increase as $\omega_{1}$.
However, for $\omega_{1} \simge 0.2$, the behavior is reversed;
all the three quantities turn to decrease as in Fig. \ref{fig:Om0=0.5,lambda=0.8}.
Therefore in this intermediate region, as the value of $\omega_{1}$ increases,
the effective interaction between fermions is enhanced
for small values of $\omega_{1}$, however,
turns to be weakened for large values of $\omega_{1}$.

This crossover is closely related with complete softening of the boson field.
Figure \ref{fig:Om0=0.5,lambda=0.4,Om*} shows
the effective frequency of the boson field $\omega^{*}$
which is given by a pole of the Green's function for boson as
\begin{equation}
\label{eq:omega*}
\omega^{*} = \sqrt{(\omega_{0}-\omega_{1})^{2} + \Pi(i\omega_{n}=0)},
\end{equation}
where $\Pi$ is the self-energy for boson.
The frequency $\omega^{*}$ goes to zero at the value of $\omega_{1}$
where the crossover from the enhancement to the weakening of the effective interaction
exhibited in Fig. \ref{fig:Om0=0.5,lambda=0.4}.

\subsection{Phase Diagram}
\label{SubSec:Phase Diagram}

We systematically investigate the crossover found in the previous section
changing the parameter $\omega_{0}$ and $U$.
Figure \ref{fig:Om0=0.5,2.0,Om*} shows
the values of $\omega^{*}$ as a function of $U$ for the cases of
(a) $\omega_{0} = 0.5$ and (b) $\omega_{0} = 2.0$.
For finite values of the width $\omega_{1}$,
the frequency $\omega^{*}$ goes to zero in both cases.
We determine the crossover values of $U$
by this complete softening of $\omega^{*}$
for $\omega_{0}$ and $\omega_{1}$.

Figure \ref{fig:phase diagram} summarizes the phase diagram
for the crossovers determined by the above criterion.
This indicates  the boundary
between the weak-fluctuation and the strong-fluctuation regimes
as discussed in Sec. \ref{Sec:Discussions}.
The most important point in this phase diagram is that
especially for large $\omega_{0}$,
the energy scale of $U$ for this crossover is quite different from $U^{*}$
determined in Sec. \ref{SubSec:Dispersionless Boson}.
This suggests that there is another parameter which controls the
onset of  strong fluctuations as discussed
in the next section \ref{Sec:Discussions}.

\section{Discussion}
\label{Sec:Discussions}

In this section, we discuss the results obtained in Sec. \ref{Sec:Results}.
Perturbative arguments are applied to discuss
the enhancement of the effective interaction between fermions
in the weak-coupling
\cite{Migdal1958}
and the atomic regions.
\cite{Holstein1959a,Holstein1959b,Mahan1981}
In the strong coupling regime away from the anti-adiabatic limit,
the weakening of the effective interaction and the instability of the small-polaron state
are discussed as a consequence of the strong fluctuations of the boson fields
accompanied by complete softening.
The phase diagram is examined to clarify the  parameters 
which control the onset of the strong fluctuations.

In the weak coupling region,
the dispersion  width $\omega_{1}$ enhances
the effective interaction between fermions in our DMF solutions.
The absolute value of the self-energy for fermion is enhanced.
A first-order perturbation in the coupling parameter $\lambda$
concludes that the self-energy
$\Sigma$ becomes larger as the width $\omega_{1}$ increases
since ${\cal D}_{0}$ increases as $\omega_{1}$.
\cite{Migdal1958}
Thus, perturbation theory suggests that
the boson dispersion increases the effective interaction between fermions
in the weak coupling limit.
This enhancement can be understood intuitively as follows:
In the weak coupling region, the density of states
for both fermion and boson are not altered drastically by the fermion-boson interaction;
a rigid-band picture should be justified.
For a finite $\omega_{1}$, the band edge of the boson density of states is lowered linearly.
Then, the effective interaction (\ref{eq:U_eff}) is mainly mediated
by the bosons near the band edge as
\begin{equation}
U_{\rm eff} (i \omega_{n}) \sim 
\frac{\lambda^{2}}{(i \omega_{n})^{2} - (\omega_{0} - \omega_{1})^{2}}.
\end{equation}
Thus the absolute value $|U_{\rm eff}|$ becomes large as the value of $\omega_{1}$ increases.
Therefore, the enhancement of the effective interaction in the weak coupling region
can be understood as the decrease  of the effective boson frequency.
Our results in Sec. \ref{SubSec:Weak Coupling Limit}
are consistent with this perturbative argument.

We now turn to the atomic  limit,  $W \ll \omega_{0}$ and $U$.
Now the fermion hopping term in Eq. (\ref{eq:H_F})
is a perturbation to the  terms (\ref{eq:H_B}) and (\ref{eq:H_I}).
If we apply the canonical transformation to diagonalize the unperturbed terms
according to the small-polaron theory,
\cite{Holstein1959a,Holstein1959b,Mahan1981}
we obtain the expression of the Hamiltonian as
\begin{eqnarray}
{\cal H} &=& \sum_{\mbox{\boldmath $q$}} \omega_{\mbox{\boldmath $q$}}
\left( a_{\mbox{\boldmath $q$}}^{\dagger}
a_{\mbox{\boldmath $q$}} + \frac{1}{2} \right)
- \sum_{i \alpha} c_{i \alpha}^{\dagger} c_{i \alpha} \Delta
\nonumber
\\
&+& \sum_{ij, \alpha} t_{ij} c_{i \alpha}^{\dagger} c_{j \alpha} X_{i}^{\dagger} X_{j},
\label{eq:H_SP}
\end{eqnarray}
where $\Delta$ is the stabilization energy of polarons given by
\begin{equation}
\Delta = \sum_{\mbox{\boldmath $q$}} \frac{{\lambda}^{2}}
{\omega_{\mbox{\boldmath $q$}}},
\label{eq:Delta}
\end{equation}
and the operator $X_{i}$ takes the form
\begin{equation}
X_{i} = \exp \Biggl[ \ \sum_{\mbox{\boldmath $q$}}
e^{{\rm i} \mbox{\boldmath $q$} \cdot \mbox{\boldmath $r$}_{i}}
\frac{\lambda}{\omega_{\mbox{\boldmath $q$}}}
(a_{\mbox{\boldmath $q$}} - a_{-\mbox{\boldmath $q$}}^{\dagger}) \ \Biggr].
\end{equation}
The third term of the Hamiltonian (\ref{eq:H_SP}) indicates that
the hopping occurs not as a bare fermion but as a combined object between fermion and boson.
Each fermion is associated by a local boson.
This is called the small-polaron state.

In the unperturbed state ($t_{ij} = 0$), the polarons are almost localized in real space
with the stabilization energy $\Delta$ given by Eq. (\ref{eq:Delta}).
When one increases the width of the boson dispersion $\omega_{1}$,
the stabilization energy $\Delta$ increases.
This makes the polarons more strongly localized.
The delocalization of the polarons is a second-order perturbation in $t$-term
in Eq. (\ref{eq:H_SP}) since the polarons consist of the double occupancy of fermions in this model.
The hopping matrix by the second-order process is suppressed by $\omega_{1}$
because the intermediate state in the perturbation costs the energy $\Delta$.
The operator $X_{i}$ does not change the result in this limit of $\omega_{0} \gg W$.
Therefore a finite width of the boson dispersion suppresses
the band motion of the polarons 
and enhances the effective interaction in this limit.
Our results in Sec. \ref{SubSec:Atomic Limit}
are in good agreement with this argument based on the small-polaron theory.

We now discuss the strong coupling regime away from the anti-adiabatic limit
studied in Sec. \ref{SubSec:Strong Fluctuation Regime}.
In this regime, the polaron state is formed by the strong coupling, however
the boson fields are loosely bound to the fermions due to a finite $\omega_{0}$,
compared to the previous atomic limit with $\omega_{0} \gg W$
where the boson fields react instantaneously to fermion motions.
This leads to large fluctuations in the boson fields by the hopping of fermions.
Thus this regime is characterized by these strong fluctuations,
which is the reason why any perturbation cannot be applied.
These fluctuations may in turn accelerate the delocalization of fermions
through the mutual feedback effects of the fermion-boson coupling.

A finite width of the boson dispersion $\omega_{1}$ introduced in our calculations
increases the fluctuations of the boson fields.
The bosons are not localized and gain their kinetic energy through the dispersion.
By tuning the width $\omega_{1}$,
we can control the fluctuations of the boson fields by hand.
Our results in Sec. \ref{SubSec:Strong Fluctuation Regime} clearly exhibited that
the effective interaction between fermions is weakened by $\omega_{1}$.
This is considered to be a consequence of the strong fluctuations of the boson fields
enhanced by $\omega_{1}$ which tend to make fermions more delocalized.
This behavior is elucidated for the first time
by our method which fully includes the mutual feedback in many-body systems.

\begin{figure}
\epsfxsize=7cm
\centerline{\epsfbox{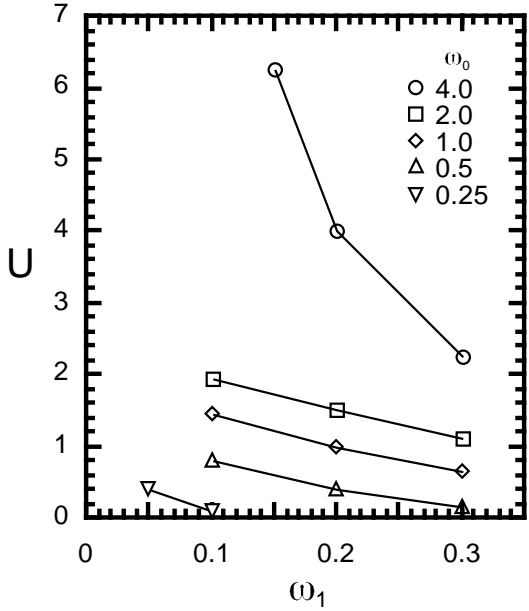}}
\caption{
The phase diagram of the crossover for various values of $\omega_{0}$.
The boundaries are determined by the complete softening of
the effective boson frequency.
The lines are guides to eye.
}
\label{fig:phase diagram}
\end{figure}

In the intermediate coupling region,
a sharp crossover was found by changing the value of $\omega_{1}$
in Sec. \ref{SubSec:Strong Fluctuation Regime}.
For small $\omega_{1}$, the effective interaction is enhanced by $\omega_{1}$.
Since the fluctuations are small there,
this may be smoothly connected to the behavior discussed
in the weak-coupling or the atomic regions.
The effective frequency $\omega^{*}$ defined by Eq. (\ref{eq:omega*})
becomes small but remains finite as in the perturbative regime
although the reduction of $\omega^{*}$ is large and nonlinear
in this nonperturbative regime.
When the value of $\omega_{1}$ becomes large enough
to soften the boson field completely ($\omega^{*} \rightarrow 0$),
fluctuations play a crucial role to enhance the delocalization of fermions.
The boundaries in the phase diagram in Fig. \ref{fig:phase diagram}
are the crossovers between the weak-fluctuation and
the strong-fluctuation regimes.

In the dispersionless case in Sec. \ref{SubSec:Dispersionless Boson},
we have found another crossover by the formation of the small polaron
which is, for instance, characterized by the development of the double-peak structure
in the probability $P(x)$.
The critical value of $U$ for this crossover, $U^{*}$, changes
from $U^{*} \sim 3$ in the limit of $\omega_{0} \gg W$ (anti-adiabatic limit)
to $U^{*} \sim 1$ in the limit of $\omega_{0} \ll W$ (adiabatic limit)
as shown in the schematic phase diagram in the plane ($1/\omega_{0}, U$)
in Fig. \ref{fig:phase diagram,Om1=0} (dotted gray line).
On the other hand,
the crossover to the strong fluctuation regime in Fig. \ref{fig:phase diagram}
appears at much larger values of $U$ than $U^{*}$
especially in the anti-adiabatic regime with large but finite $\omega_{0}$.
This strongly suggests the importance of another energy scale
to characterize the strong fluctuation regime
which is not clearly found in the dispersionless case.

The importance of such characteristic energy scale has been pointed out
also in a  previous mean-field study.
\cite{Feinberg1990}
The parameter is defined by the ratio of the fermion-boson interaction
to the spring constant of boson fields, $\eta = \lambda / M \omega_{0}^{2}$.
In the case of $\eta < 1$, since the fermion-boson interaction is weak
compared to the stored energy in the boson field,
the single-boson process should be important.
In the case of $\eta > 1$, the fermion-boson interaction is strong enough
to excite a large   numbers of bosons, i.e., 
multiboson processes become important.
In the previous study,
\cite{Feinberg1990}
the importance of fluctuations of the boson fields has been suggested
in the latter multiboson regime.

\begin{figure}
\epsfxsize=7cm
\centerline{\epsfbox{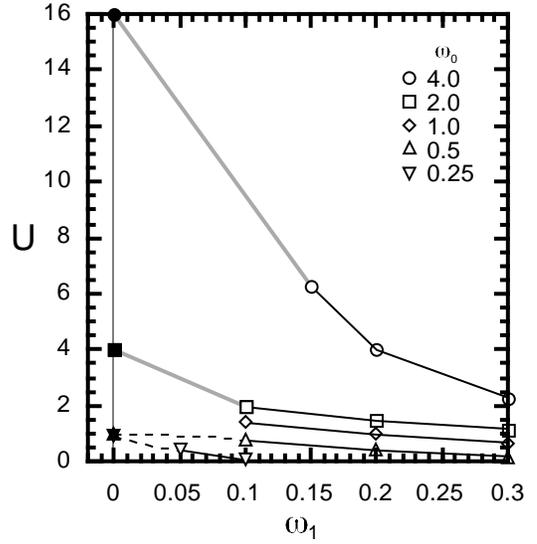}}
\caption{
The values of $U$ and $\omega_{1}$ at which the crossover from
the weak-fluctuation to the strong-fluctuation regime takes place.
Notice that these values extrapolate smoothly to the values of $U$ at $\eta=1$
for $\omega_{0} = 4$ and $2$ (anti-adiabatic regime) and to
the $U^{*}$ for $\omega_{0} = 0.5$ and $0.25$ (adiabatic regime)
in the limit of $\omega_{1} = 0$.
See the text for details.
}
\label{fig:phase diagram2}
\end{figure}

In the plane ($1/\omega_{0}, U$),
the crossover between the single-boson and the multiboson regimes
occurs at $\eta = 1$,
i.e., for  $U = \omega_{0}^{2}$ and is  shown in Fig. \ref{fig:phase diagram,Om1=0} (solid line).
The line of $\eta = 1$ becomes much larger than $U^{*}$ in the anti-adiabatic regime.
If we plot these values of $U(\eta=1)$ on the axis of $\omega_{1} = 0$
in Fig. \ref{fig:phase diagram},
the crossover boundaries seem to be smoothly connected to these values in the anti-adiabatic region.
We demonstrate this behavior in Fig. \ref{fig:phase diagram2}
for $\omega_{0} = 4$ and $2$ (gray lines).
This indicates that in the anti-adiabatic regime,
the line of $\eta = 1$ corresponds to the crossover
between the weak-fluctuation and the strong fluctuation regimes
in the dispersionless case.

On the other hand, in the adiabatic regime with small but finite $\omega_{0}$,
the value of $U$ for the boundary in Fig. \ref{fig:phase diagram} is
not smoothly connected to the value of $U$ for $\eta = 1$.
For instance, in the case of $\omega_{0} = 0.5$,
$\eta=1$ gives $U=1/4$ which is much smaller than the boundary.
This suggests that the condition of $\eta > 1$ does not characterize
the strong fluctuation regime in this adiabatic region.

To understand this behavior in the adiabatic regime,
let us discuss the adiabatic limit of $\omega_{0} \rightarrow 0$.
In this limit, the boson fields behave as classical fields
which do not fluctuate in the imaginary time direction.
Boson fluctuations only come from the fluctuation of the value of the field $x$
which is constant in time.
Then even if the system is in the region of $\eta > 1$,
the fluctuations of the boson fields are small when $U$ is smaller than $U^{*}$
since the fields feel a deep single-well potential
as indicated in the probability $P(x)$.
The boson fields begin to fluctuate when $U$ becomes comparable to $U^{*}$
where the potential for $x$ softens around $x=0$.
Therefore in the adiabatic limit,
the strong fluctuation regime should be characterized not by $\eta > 1$ but by $U > U^{*}$.

Based on this argument, if we plot the value of $U^{*}$ on the axis of $\omega_{1}=0$,
the crossover boundaries in the adiabatic regime
seem to be smoothly connected to these values
for the cases of $\omega_{0} = 0.5$ and $0.25$
 in Fig. \ref{fig:phase diagram2} (dotted lines).

These results reveal that
the boson fluctuations which are strong enough to delocalize fermions appear in a different way
in the anti-adiabatic and the adiabatic regimes.
In the anti-adiabatic regime ($\omega_{0} > W$), 
the small-polaron state is formed at $U \simge U^{*}$.
In the region with $U > U^{*}$ and $\eta < 1$, however,
the fluctuations of the boson fields are small
in the sense that the single-boson process contributes mainly and
that the effective boson frequency is finite.
If $\eta$ becomes larger than $1$,
the boson field is softened and
the boson fluctuations play a crucial role through the multiboson process.
On the other hand, in the adiabatic regime ($\omega_{0} < W$),
the fluctuations do not become large until $U \sim U^{*}$ even if $\eta$ is larger than $1$.
The fluctuations are mainly the classical origin there.

To  summarize, the strong fluctuations of the boson fields become important
only when the conditions $U > U^{*}$ and $\eta > 1$ are both satisfied.
These conditions are shown as the hatched area in Fig. \ref{fig:phase diagram,Om1=0}.
This area is strongly-correlated region for both fermions and bosons.
The criterion for the formation of small polarons, $U \sim U^{*}$,
corresponds to a competition between the kinetic energy of fermions $W$ and
the effective interaction $U$.
The line of $U^{*}$ may be modified according to a specific form of
the fermion-boson coupling.
On the other hand, the criterion $\eta \sim 1$ corresponds to
a competition between the stored energy of the boson field and
the coupling energy to fermions.
Thus the hatched area in Fig. \ref{fig:phase diagram,Om1=0} is
the region where correlations become strong in both standpoints
of fermions and bosons.

A subtle problem remains open about the boson softening in the dispersionless case.
As shown in Fig. \ref{fig:Om0=0.5,2.0,Om*},
when $\omega_{1}$ is zero, the effective frequency $\omega^{*}$ becomes
very small but remains finite for large values of $U$.
We note that there are finite-temperature effects;
$\omega^{*}$ is suppressed more strongly for lower temperatures.
Unfortunately we cannot conclude in this study
whether $\omega^{*}$ goes to zero even when $\omega_{1} = 0$.
In this dispersionless case, the boson density of states is a delta function
at $\varepsilon = \omega_{0}$, which is special
since the shape of the boson density of states at the bottom is important
as mentioned in Sec. \ref{SubSec:Model}.
For instance, the step-like singularity in the two-dimensional density of states
might prevent the boson field from complete softening at finite temperatures.
Though further studies are necessary for the property of DMF equations
for various types of density of states, we believe
from the results in Figs. \ref{fig:phase diagram,Om1=0} and \ref{fig:phase diagram2}
that the nonlinear suppression of $\omega^{*}$ is relevant to strong fluctuations
of bosons and that there are two important energy scales even when $\omega_{1} = 0$.

\section{Summary and Concluding Remarks}
\label{Sec:Summary}

We have investigated the effects of the boson dispersion
in a system of dynamical mean-field equations describing
coupled fermion-boson systems.
The analysis of the equations revealed that the boson dispersion
plays a crucial role in a wide region of parameters.
By introducing a parameter for the width of the dispersion in the model,
we can control the fluctuations of the boson fields.
To handle the boson fluctuations and the feedback effects,
we have extended the dynamical mean-field theory
to determine the Green's functions for both fermion and boson in the self-consistent way.
In the ordinary framework for the dispersionless case,
the channel for the boson Green's function is frozen
in the sense that the bare impurity Green's function is fixed and
unrenormalized from the noninteracting one.
The renormalization of the bare impurity Green's function for boson is very important
since the bare impurity Green's function is directly related to
the effective interaction between fermions.
The equations in the extended dynamical mean-field theory
are solved by using quantum Monte Carlo technique.

The main result in the models with dispersive bosons is that
in the strong coupling regime away from the anti-adiabatic limit,
the fluctuations of the boson fields become relevant to accelerate the delocalization of fermions.
The effective interaction between fermions is weakened
as the width of the boson dispersion increases  in this regime.
This behavior is explicitly shown for the first time by our method
which fully includes the mutual feedback effects.
The crossover to this nonperturbative regime is closely correlated
with softening of the boson field.
We have examined
the  phase diagram where this  strong fluctuation occurs
by tuning  the coupling parameter and the width of the dispersion.
The strong fluctuations to delocalize fermions become relevant
when the small-polaron state is formed and the multiboson processes become important.
The small polarons become stable when the effective interaction between fermions
overcomes the fermion band energy.
The multiboson regime is characterized by a coupling parameter larger than the boson energy.
Thus the strong fluctuation regime is the strong correlated region
for both fermions and bosons.
As the coupling parameter increases, the boson fluctuations appear in a different way
between in the adiabatic and the anti-adiabatic regimes.
In the adiabatic regime, the fluctuations are mainly classical
which are enhanced by the softening of the potential for the boson fields
in the formation of the small-polaron state.
On the other hand, in the anti-adiabatic regime,
the small polarons are formed in the single-boson regime,
where the dynamical fluctuations are small and
the effective boson frequency is finite.
The boson fluctuations do not play a crucial role
until the system enters in the multiboson regime
by complete softening of bosons.

The onset of the strong fluctuations occurs near the region
where the boson degrees of freedom soften.
In this paper we have studied the DMF equations
in the absence of any freezing of the boson degrees of freedom.
These effects together with generalizations to states with different symmetries
and other generalizations are currently under investigation.

Our results imply that the behavior of the boson fluctuations may depend on
the specific form of the boson density of states.
Different forms of the density of states should be tested
in the present DMF framework in a future study.
Especially we are interested in the two-dimensional case
with a step-like singularity at the edge which might
be free from complete softening at finite temperatures.
Boson fluctuations in this case may lead to light-mass bipolaronic states,
which would give some insights into the high-temperature superconductivity
in Cu-oxide materials where fermions strongly couple with spin fluctuations.
We plan to understand this two-dimensional case in a later publication.

The dynamical mean-field equations allow us to vary
the width of the boson dispersion in the calculations.
This reveals the interesting properties in the strong fluctuation regime.
The tuning of the electronic bandwidth  has been the subject
of a great deal of  
theoretical and experimental work
\cite{Imada1998}.
Our work suggests the possible interest of varying the 
boson dispersion experimentally even though  this may
be easier in systems where the bosons are spin fluctuations
whose dispersion determined by
exchange interactions can be controlled more easily
than optical phonon dispersions. Another possibility may
be the realization of 
the dynamical mean-field theory in a random model.   

There are many materials which satisfy the above conditions for the strong fluctuation regime.
In many physical situations, the fermion bandwidth $W$ is large or comparable to $\omega_{0}$,
which makes possible to access to the strong fluctuation regime by a relatively weak coupling.
Our method provides a powerful theoretical tool to examine the physical properties in this regime.
We can apply it to more realistic models
including orbital degrees of freedom of electrons,
different normal modes of phonons, or interactions between fermions.
Such extensions are now under investigation.

\section*{Acknowledgement}

Y. M. acknowledges the financial support of
Research Fellowships of Japan Society for the Promotion
of Science for Young Scientists.
G. K. is supported by the NSF under DMR  95-29138. 



\begin{figure}
\epsfxsize=8cm
\centerline{\epsfbox{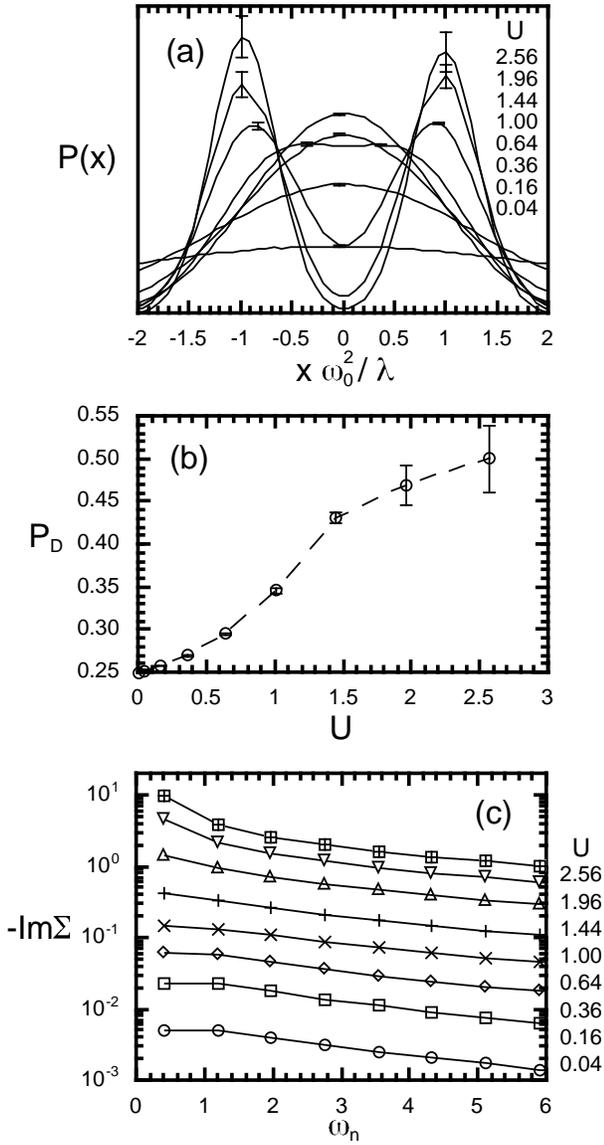}}
\caption{
Results for the dispersionless model with $\omega_{0} = 0.5$ at $\beta=8$;
(a) the probability function of the boson fields $x$,
(b) the probability of the double occupancy, and
(c) the imaginary part of the self-energy for fermion as a function of Matsubara frequency.
In (a), the typical errorbars are shown at the peaks of the distributions.
The lines in (b) and (c) are guides to eye.
}
\label{fig:Om0=0.5,Om1=0}
\end{figure}

\begin{figure}
\epsfxsize=8cm
\centerline{\epsfbox{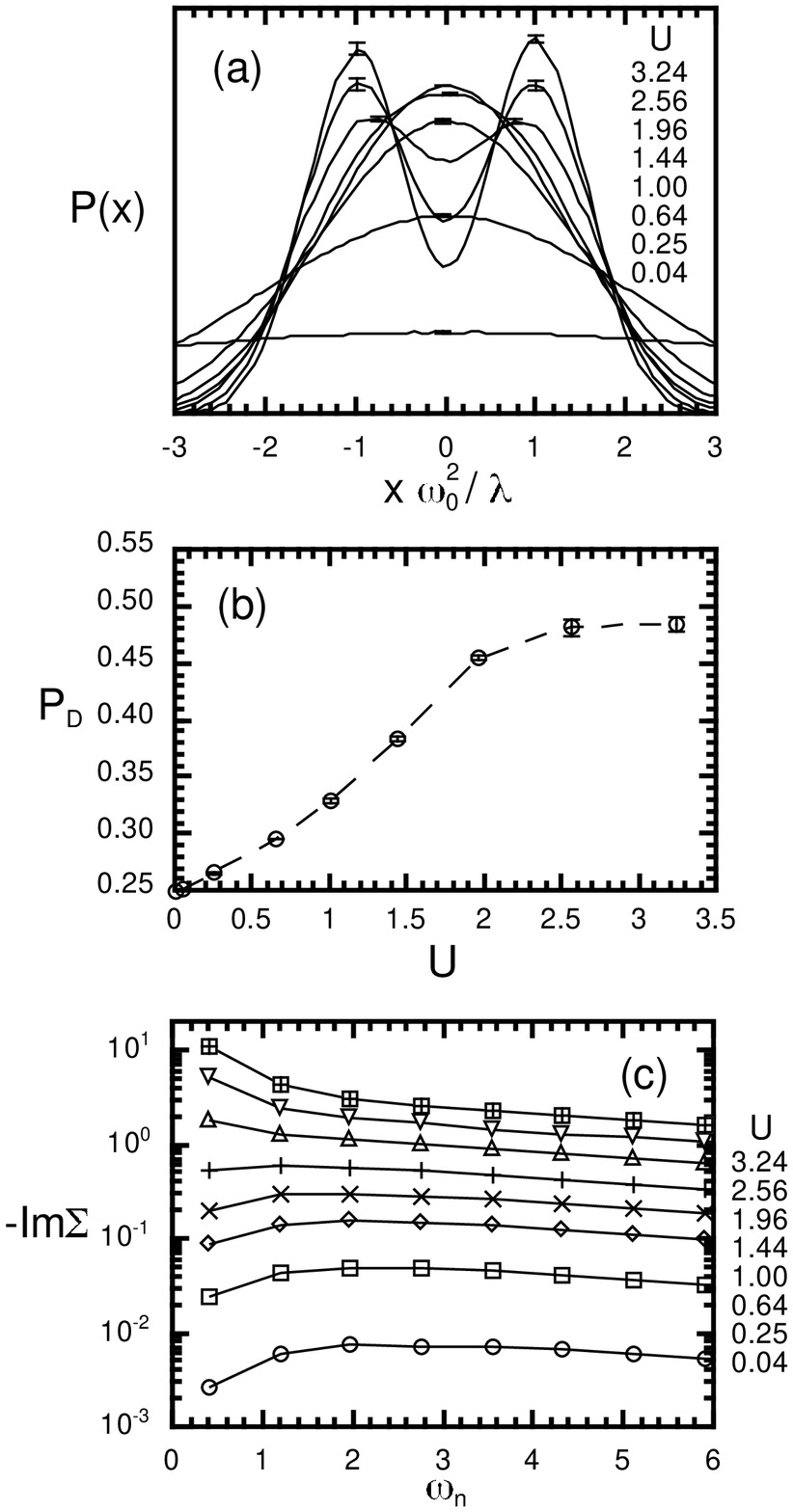}}
\caption{
Results for the dispersionless model with $\omega_{0} = 2$.
}
\label{fig:Om0=2.0,Om1=0}
\end{figure}

\begin{figure}
\epsfxsize=8cm
\centerline{\epsfbox{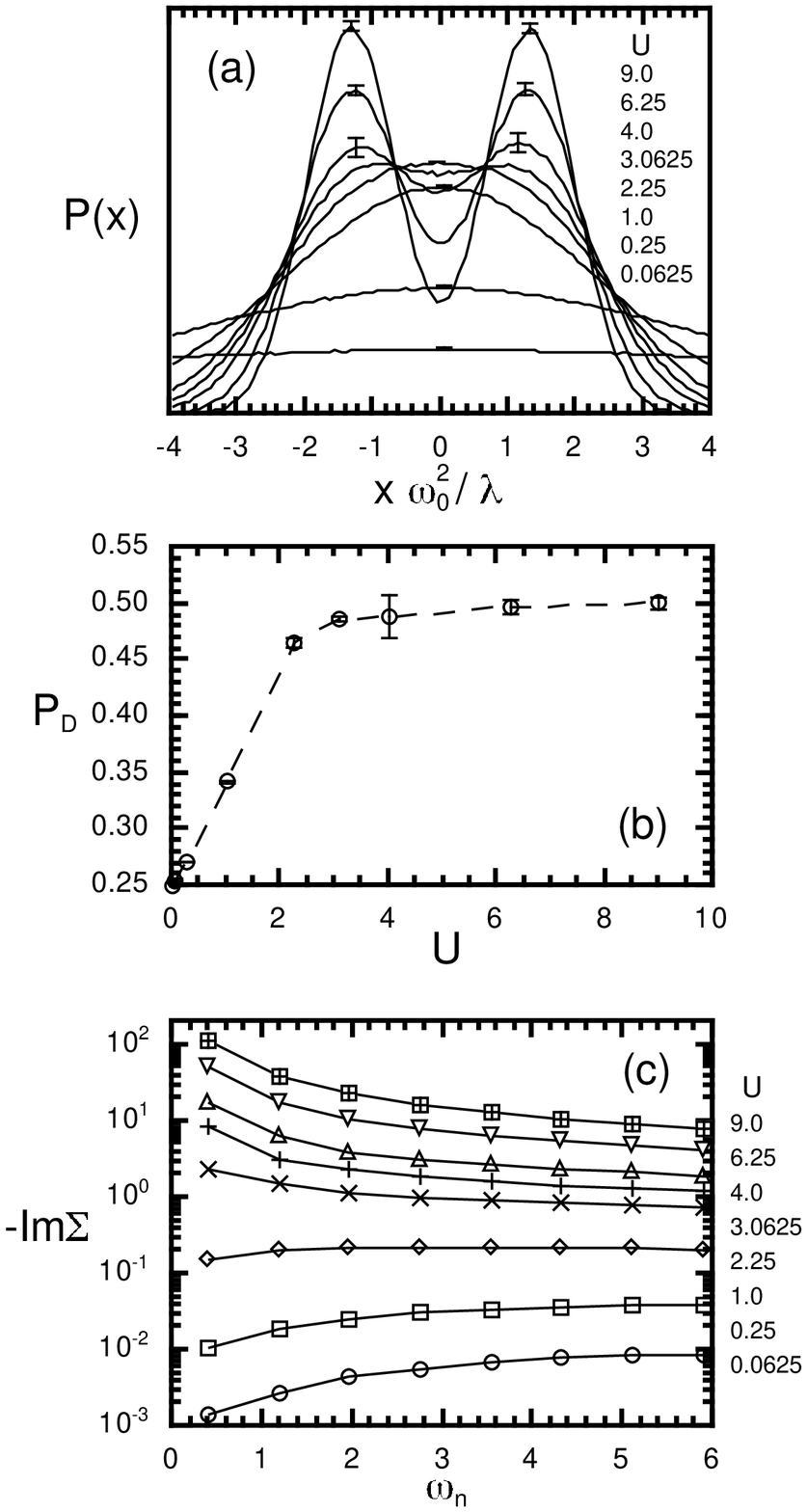}}
\caption{
Results for the dispersionless model with $\omega_{0} = 8$.
}
\label{fig:Om0=8.0,Om1=0}
\end{figure}

\begin{figure}
\epsfxsize=8cm
\centerline{\epsfbox{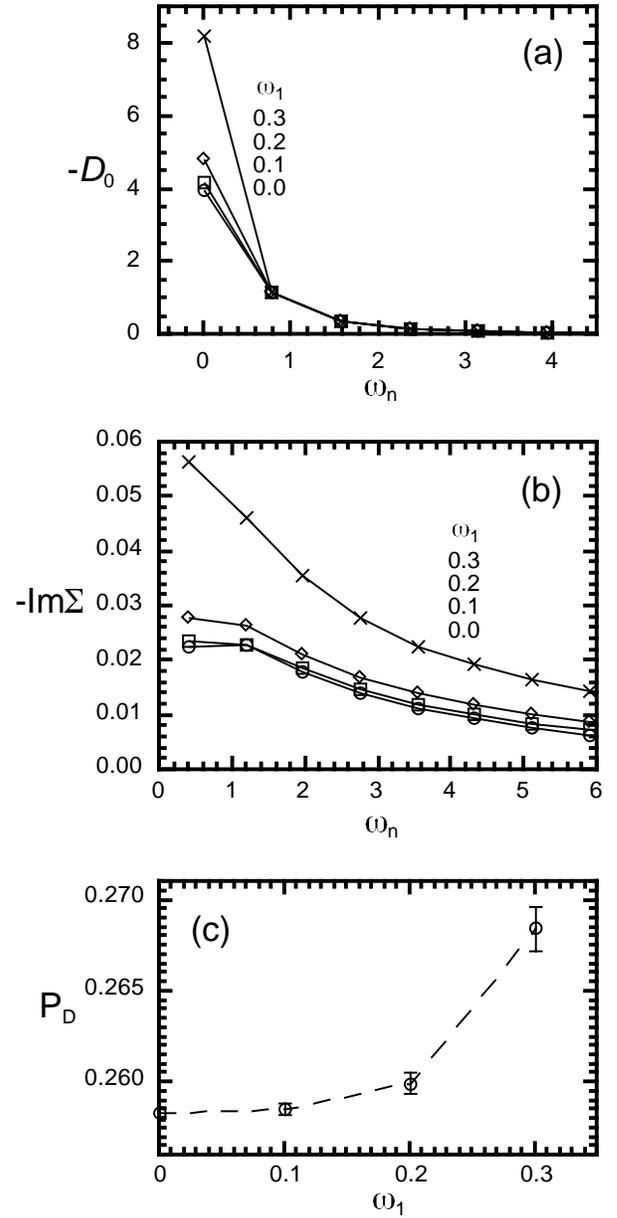}}
\caption{
Results for the dispersive boson model in the weak coupling regime
with $\omega_{0} = 0.5$ and $U = 0.16$ ($\lambda = 0.2$) at $\beta=8$;
(a) the bare impurity Green's function for boson as a function of Matsubara frequency,
(b) the imaginary part of the self-energy for fermion, and
(c) the probability of the double occupancy as a function of the width of the boson dispersion.
The lines are guides to eye.
}
\label{fig:Om0=0.5,lambda=0.2}
\end{figure}

\begin{figure}
\epsfxsize=8cm
\centerline{\epsfbox{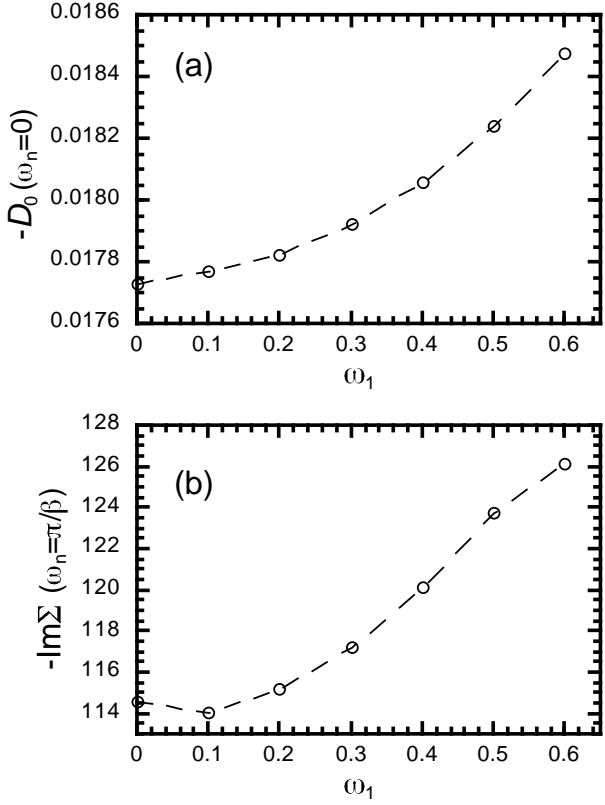}}
\caption{
Results for the dispersive boson model in the atomic regime
with $\omega_{0} = 8$ and $U = 9$ ($\lambda = 24$) at $\beta=8$;
(a) the bare impurity Green's function for boson at zero Matsubara frequency and
(b) the imaginary part of the self-energy for fermion at the smallest Matsubara frequency.
The lines are guides to eye.
}
\label{fig:Om0=8,lambda=24}
\end{figure}

\begin{figure}
\epsfxsize=7cm
\centerline{\epsfbox{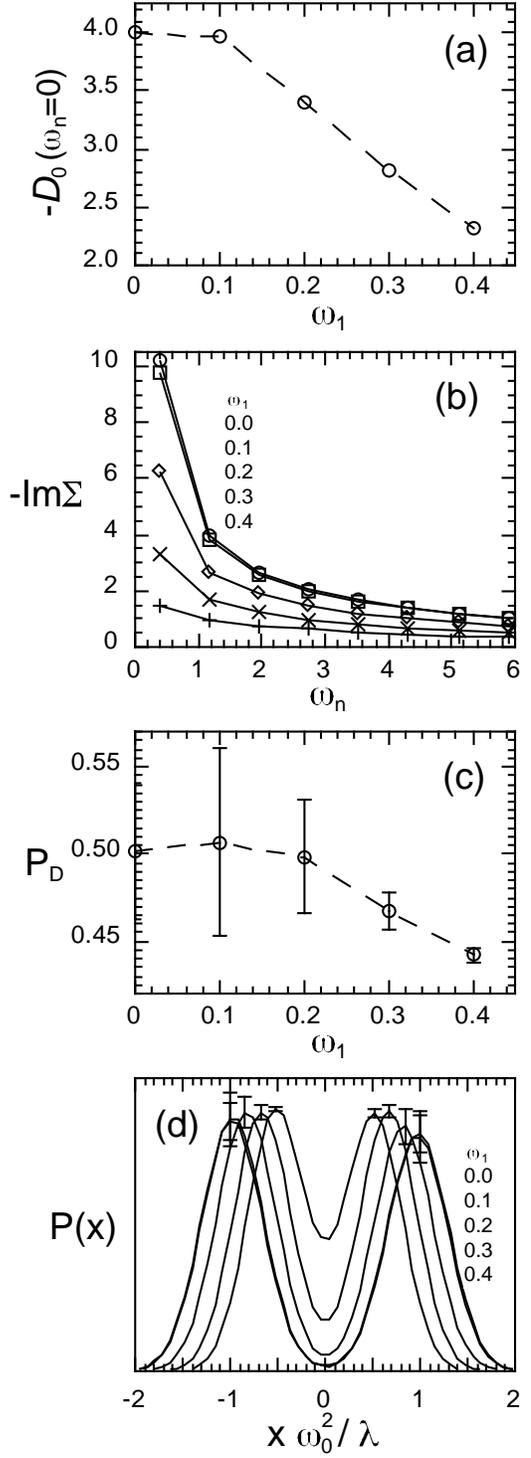}}
\caption{
Results for the dispersive boson model
with $\omega_{0} = 0.5$ and $U = 2.56$ ($\lambda = 0.8$) at $\beta=8$;
(a) the bare impurity Green's function for boson at zero Matsubara frequency,
(b) the imaginary part of the self-energy for fermion,
(c) the probability of the double occupancy, and
(d) the probability function of the boson fields $x$.
The lines in (a)-(c) are guides to eye.
In (d), the typical errorbars are shown at the peaks of the distributions.
}
\label{fig:Om0=0.5,lambda=0.8}
\end{figure}

\begin{figure}
\epsfxsize=8cm
\centerline{\epsfbox{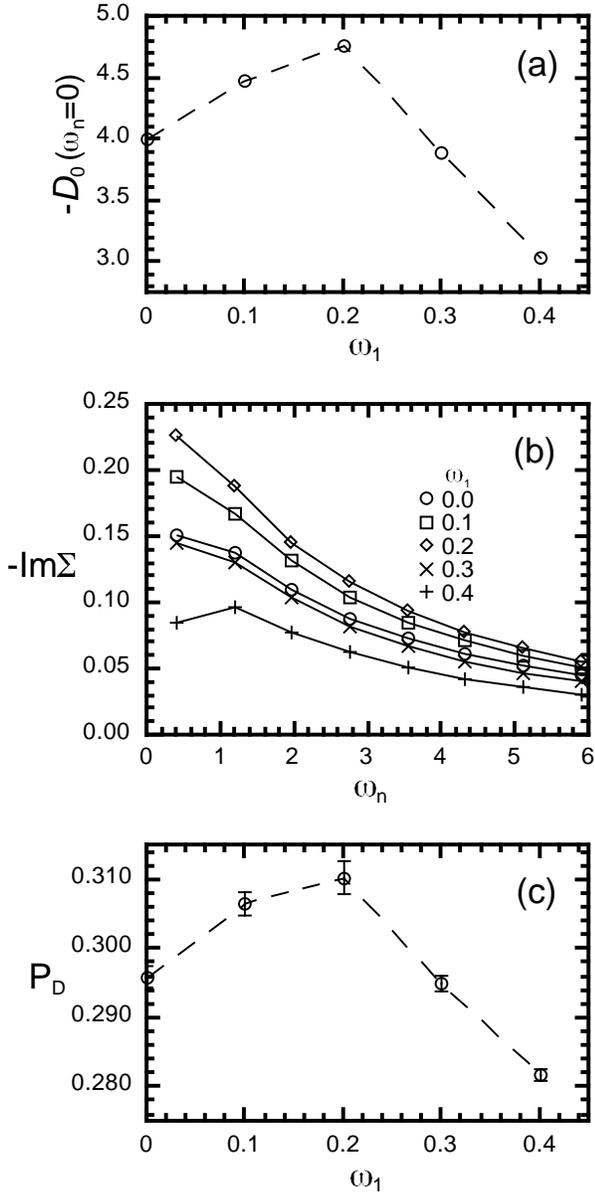}}
\caption{
Crossover in the intermediate coupling region
with $\omega_{0} = 0.5$ and $U = 0.64$ ($\lambda = 0.4$) at $\beta=8$;
(a) the bare impurity Green's function for boson at zero Matsubara frequency,
(b) the imaginary part of the self-energy for fermion, and
(c) the probability of the double occupancy.
The lines are guides to eye.
}
\label{fig:Om0=0.5,lambda=0.4}
\end{figure}

\begin{figure}
\epsfxsize=8cm
\centerline{\epsfbox{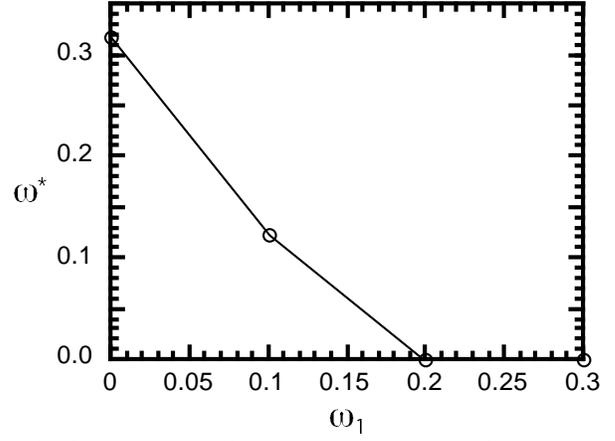}}
\caption{
The effective frequency of boson in the case of $\omega_0=0.5$ and $U=0.64$
at $\beta=8$.
The line is a guide to eye.
}
\label{fig:Om0=0.5,lambda=0.4,Om*}
\end{figure}

\begin{figure}
\epsfxsize=8cm
\centerline{\epsfbox{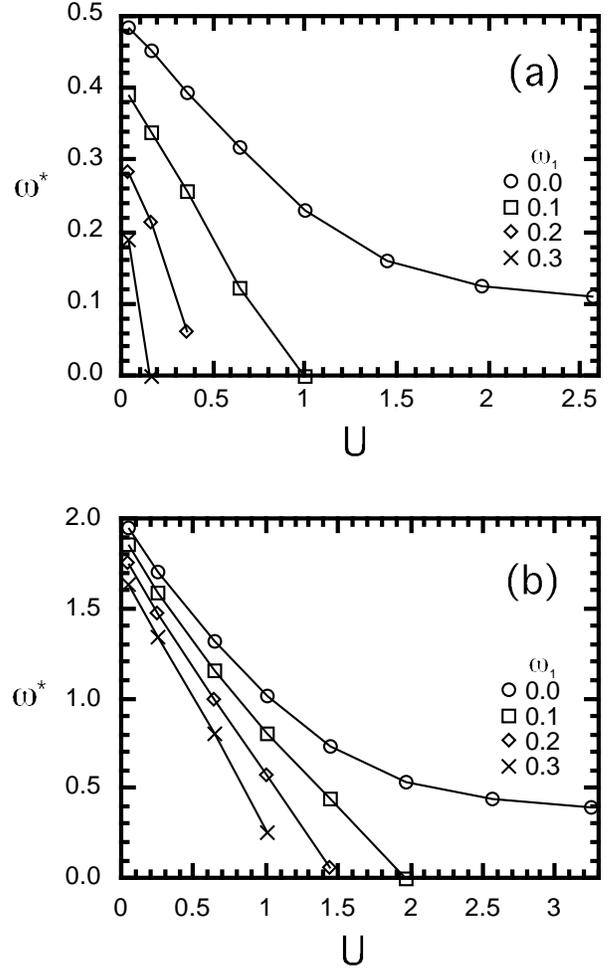}}
\caption{
The softening of the effective boson frequency for
(a) $\omega_{0}=0.5$ and (b) $\omega_{0}=2.0$ at $\beta=8$.
The lines are guides to eye.
}
\label{fig:Om0=0.5,2.0,Om*}
\end{figure}

\begin{figure}
\epsfxsize=7cm
\centerline{\epsfbox{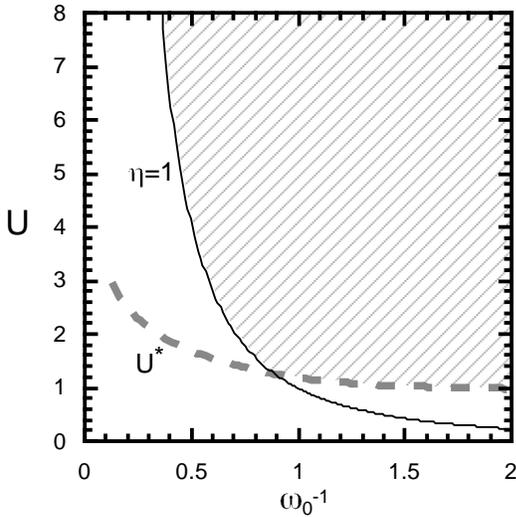}}
\caption{
Schematic phase diagram for the dispersionless case.
The critical value for the formation of the small-polaron state $U^{*}$
is shown as the dotted gray line.
The solid line indicates the
boundary  $\eta=1$.
The hatched area is the strong fluctuation region.
See the text for details.
}
\label{fig:phase diagram,Om1=0}
\end{figure}

\end{document}